\documentclass[twoside,showpacs,superscriptaddress,twocolumn,floatfix,a4paper,aps]{revtex4}

\usepackage{color}
\usepackage{graphicx}
\usepackage[utf8x]{inputenc}
\usepackage[T1]{fontenc}

\newcommand{\etal}{\textit{et al.}}

\newcommand\extra[1]{}
\begin{document}

\title{Resource-efficient linear-optical quantum router}

\author{Karel Lemr}
\email{k.lemr@upol.cz}
\affiliation{RCPTM, Joint Laboratory of Optics of Palacký University and Institute of Physics of Academy of Sciences of the Czech Republic, 17. listopadu 12, 772 07 Olomouc, Czech Republic}
\author{Karol Bartkiewicz}
\affiliation{RCPTM, Joint Laboratory of Optics of Palacký University and Institute of Physics of Academy of Sciences of the Czech Republic, 17. listopadu 12, 772 07 Olomouc, Czech Republic}
\author{Antonín Černoch}
\affiliation{RCPTM, Joint Laboratory of Optics of Palacký University and Institute of Physics of Academy of Sciences of the Czech Republic, 17. listopadu 12, 772 07 Olomouc, Czech Republic}
\author{Jan Soubusta}
\affiliation{Institute of Physics of Academy of Sciences of the Czech Republic, Joint Laboratory of Optics of PU and IP AS CR, 
   17. listopadu 50A, 779 07 Olomouc, Czech Republic}

\date{\today}

\begin{abstract}
All-linear-optical scheme for fully featured quantum router is presented. This device directs the signal photonic qubit according to the state of one control photonic qubit. In the introduction we formulate the list of requirements imposed on a fully quantum router. Then we describe our proposal showing the exact principle of operation on a linear-optical scheme. Subsequently we provide generalization of the scheme in order to optimize the success probability by means of a tunable controlled-phase gate. At the end, we show how one can modify the device to route multiple signal qubits using the same control qubit.
\end{abstract}

\pacs{42.50.Dv 03.67.Hk 03.67.Lx}

\maketitle

\section{Introduction}
Quantum communications represent very important part of a rapidly developing research area
called quantum information processing \cite{Nielsen_QCQI,Zeilinger_QIP}.
Communications networks are nowadays an indispensable technology allowing people 
to transmit information quickly over large distances. 
In classical communications networks classical physics laws are used for their construction.
The recent research in quantum physics and quantum information suggests that quantum laws 
of nature can provide significant improvement of capabilities of communications devices
\cite{Barenco95,OBrien03,Brassard05}, 
while using similar resources and keeping similar network architecture
\cite{MedhiRamasamy_Net,SalehTeich_Photonics}.  
Not surprisingly, a significant amount of theoretical and experimental research 
has been dedicated to the concept of quantum communications networks
\cite{Kimble08}. 
Most notable result of this effort is a number of protocols for quantum cryptography 
\cite{Gisin02,Scarani09}, 
which is a method for unconditionally secure transmission of information using 
various quantum properties of information carriers. 
Quantum communications networks can benefit from purely quantum effects such as 
entanglement or probabilistic nature of measurement. 
%On the other hand they share some common traits with their classical analogues such as for instance similar proposals for topology.

The more complex both the classical and quantum networks are, the more pronounced 
is the need for correct routing of the signal from its source to its intended destination 
\cite{Jackel90,Ruiqiang11}. 
Classical routers are well known ingredient of classical networks allowing to direct 
signal information according to control information (e.g. IP address) \cite{rfc1180}. 
The analogy between classical and quantum networks suggests that complex quantum 
networks would also require elaborate routing protocols.
This need is even more pronounce since in contrast to classical information, 
one can not perfectly duplicate an unknown qubit of quantum information
\cite{Buzek96}. However, approximate cloning is possible and has been intensely studied
both theoretically and
experimentally over the last decades \cite{cloning1,cloning2,cloning3}
resulting in, e.g.,
establishing and implementing optimal state-dependent cloning for a wide
class of qubit
distributions \cite{Bartkiewicz10,Lemr12}.
Nevertheless the impossibility of perfect cloning prevents using the concept of multi-directional broadcast (known from 
classical networks) in a quantum network.
\begin{figure}
\centerline{\includegraphics[scale=1]{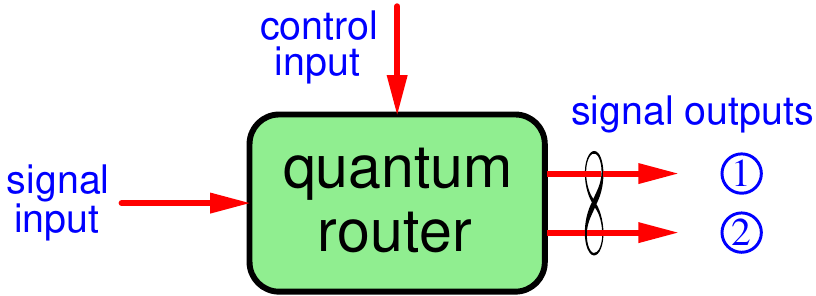}}
\caption{(Color online) Conceptual scheme of a quantum router.}
\label{fig:concept}
\end{figure}

In this paper we address the problem of designing a quantum router (see conceptional 
scheme in Fig. \ref{fig:concept}). We consider the platform of individual photons and 
linear optics because of its experimental accessibility and also because of the particular 
suitability of light for information transmission
\cite{Kok07}. 
The router has to fulfil four requirements to be suitable for quantum communications networks:   
(i) Both the signal and control information have to be stored in quantum objects (qubits), 
therefore routers using classical information to route quantum signal are considered 
only semi-quantum routers. 
(ii) The signal information is unchanged under the routing operation, the degree 
of freedom used to store the signal qubit information has to be kept undisturbed. 
(iii) The router has to be able to route the signal into a coherent superposition 
of both output modes. 
(iv) The router has to work without any need for post-selection on the signal output. If the router is probabilistic, successful operation can be identified by detection on the control state.
(v) To optimize the resources of the quantum network, only individual control qubit 
is required to direct one signal qubit.

There has been several schemes for quantum router already proposed, 
some of them habe been experimentally implemented, but
none of them however meets all the five requirements defined in the above paragraph. 
The first group of these proposals uses light-matter interaction in order to achieve 
quantum routing
\cite{Zueco09,Aoki09,Hoi11}. 
Such interaction is however often very challenging for experimental implementation. 
The second group of the proposals considers solely the platform of optical interactions 
to accomplish the routing. There are some proposals for semi-quantum router, 
where the control information is classical using intensive light pulses
\cite{Hall11}. 
There are also proposals where the control information is quantum, but the signal 
state collapses in the router so the requirement (ii) is not met
\cite{chang12router}. 
Recently we have proposed a fully quantum router using only linear optics 
\cite{Lemr12_router}. 
This device however does not meet the requirement (v) since it requires two quantum 
bits to control routing of one single qubit of signal information.

% -----------------------------------------------------------------------------

\section{Principle of operation}
\begin{figure}
\centerline{\includegraphics[scale=1.1]{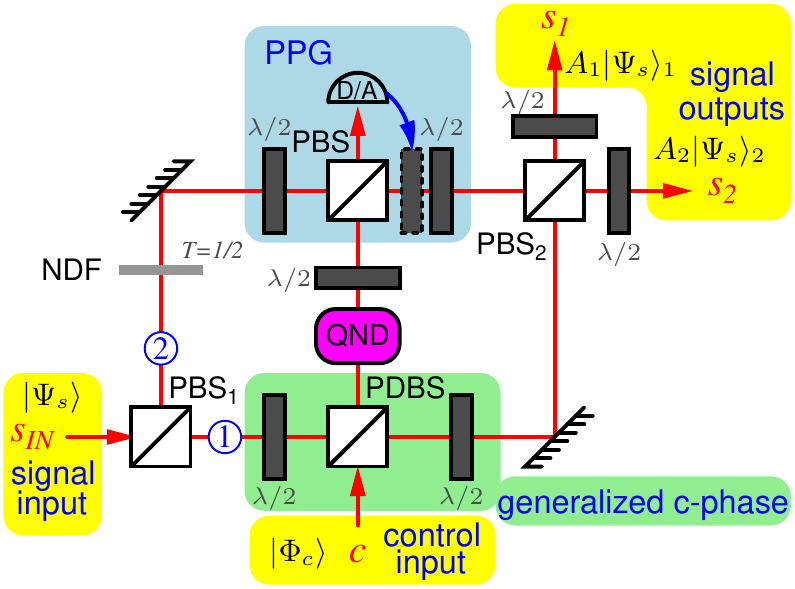}}
\caption{(Color online) Scheme of the linear-optical implementation of quantum router. PBS -- polarizing beam splitter, PDBS -- polarization dependent beam splitter, $\lambda/2$ and $\lambda/4$ -- half and quarter wave plate, NDF -- neutral density filter, c-phase -- controlled phase gate, QND -- quantum non-demolition detector, PPG -- programmable-phase gate, D/A -- polarization analysis (For more details see Ref. \cite{halenkova2012appl}).}
\label{fig:scheme}
\end{figure}

In this section we describe an all-linear-optical quantum router that meets all the five requirements mentioned in the introduction. 
The router makes use of three quantum gates: controlled-phase gate (c-phase gate)
\cite{Eisert00,Pittman03,Kieling10,Lemr_PRL106,Lemr_PRA86}, 
quantum non-demolition presence detection gate (QND gate)
%\cite{Brune90,Grangier98,Kok02,Bula13} 
\cite{Bula13} 
and programmable-phase gate (PPG)
\cite{Vidal02,Micuda08,Mikova12}. In order for the setup to work completely without need for signal post-selection, both the QND and the detector D have to be equipped with photon-number resolving detectors. Without them, the router still works, but does not fulfil the requirement (iv).
Linear-optical schemes of all of these gates are already published and with the exception of the QND gate also already tested experimentally. The reader is encouraged to get more details in the cited papers. Fig. \ref{fig:scheme} shows the scheme for linear-optical implementation of the quantum router. We propose using two degrees of freedom of individual photons: (1) polarization encoding is used to store both the signal and control qubits, (2) path (spatial mode) encoding is used for the routing operation, the signal photon is routed into superposition of two output ports.

Signal qubit enters the setup using the input port $s_{IN}$ while the control qubit enters the router using port $c$. Let us assume the signal qubit takes the form of a general quantum polarization state
\begin{equation} \label{signalstate}
| \Psi_s \rangle = \alpha |H \rangle + \beta |V \rangle ,
\end{equation}
where $|H\rangle$ and $|V\rangle$ denote the states of horizontal and vertical linear polarizations and $|\alpha^2|+|\beta^2| = 1$. For reasons apparent later, it is suitable to parametrize the state of the control qubit by angles $\theta$ and $\vartheta$
\begin{equation}
| \Psi_{c} \rangle = \cos{\theta} |H \rangle + \mathrm{e}^{\imath \vartheta} \sin{\theta} |V \rangle .
\end{equation}

After the signal qubit enters the router, it is subjected to the polarizing beam splitter (PBS$_1$) transmitting horizontally polarized light and reflecting light of vertical polarization. At this point the state of both signal and control qubits reads
\begin{eqnarray} \label{inputstate} \nonumber
|\Psi_s \rangle \otimes |\Phi_{c1} \rangle &=& 
\alpha \cos{\theta} |H_1 H_c \rangle + \alpha \mathrm{e}^{\imath \vartheta} \sin{\theta} | H_1 V_c \rangle \\ 
&+& \beta \cos{\theta} |V_2 H_c \rangle
+ \beta \mathrm{e}^{\imath \vartheta} \sin{\theta} |V_2 V_c \rangle,
\end{eqnarray}
where indexes denote spatial modes of the qubits.

For better readability, let us now consider the evolution of both arms of the interferometer (labelled in Fig. \ref{fig:scheme} as \textcircled{1} and \textcircled{2}) separately starting with the lower arm \textcircled{1} corresponding to the first two terms in Eq. (\ref{inputstate}): The horizontally polarized signal photon undergoes a generalized Hadamard transformation on a half-wave plate (HWP) rotated by -30° \footnote{In the entire paper, rotations of wave plates are given as the angle between optical axis of the wave plate and direction of horizontal linear polarization.}
$$
| H_1 \rangle \rightarrow {1 \over 2} |H_1 \rangle - {\sqrt{3} \over 2} |V_1 \rangle .
$$
Subsequently both the signal and control qubit photons enter the c-phase gate implemented using polarization dependent beam splitter PDBS of intensity transmissivities $T_V = \frac{1}{3}$ and $T_H=1$ for vertical and horizontal polarization respectively. Post-selecting only on the cases where there is one photon in signal and one photon in control mode, the transformation performed by the c-phase gate renders the investigated first two terms of (\ref{inputstate}) to
$$
{\alpha \cos{\theta} \over 2} \left( |H_1 \rangle + |V_1 \rangle \right) |H_c \rangle + {\alpha \mathrm{e}^{\imath \vartheta} \sin{\theta} \over 2 \sqrt{3}} \left( |H_1 \rangle - |V_1 \rangle \right) |V_c \rangle.
$$
In the next step, the signal qubit undergoes Hadamard transform also using a HWP yielding the terms
$$
{\alpha \cos{\theta} \over \sqrt{2}} |H_1 H_c \rangle + {\alpha \mathrm{e}^{\imath \vartheta} \sin{\theta} \over \sqrt{6}} |V_1 V_c \rangle.
$$
As mentioned above, the c-phase gate is successful only when the signal and control qubits leave the gate by separate output ports. In order to post-select only on such cases, the control qubit has to be subjected to QND presence detection via the QND gate. The QND presence detection requires an entangled pair of photons as a resource, but these photons are of a fixed quantum state, are generated locally solely for the purposes of the QND gate and do not take other part in the quantum network. The control qubit state does not change under the QND gate, but success probability factor of $\frac{1}{2}$ is added to take into account the success probability of the QND gate. More information about the QND gate can be found in Ref. \cite{Bula13}.
In the last step, the control qubit is subjected to a half-wave plate rotated at 22.5° yielding terms
$$
{\alpha \cos{\theta} \over 2 \sqrt{2}} |H_1 \rangle (|H_c \rangle + |V_c \rangle ) +
{\alpha \mathrm{e}^{\imath \vartheta} \sin{\theta} \over 2 \sqrt{6}} |V_1 \rangle (|H_c \rangle - |V_c \rangle )
$$
and then the control qubit alone impinge on the PPG composed just of a PBS. The PPG gate heralds successful operation only if there is a control qubit with horizontal polarization at its input thus projecting the signal in the lower arm \textcircled{1} onto
\begin{equation}
\label{eq:arm1}
{\alpha \cos{\theta} \over 2 \sqrt{2}} |H_1 \rangle +
{\alpha \mathrm{e}^{\imath \vartheta} \sin{\theta} \over 2 \sqrt{6}} |V_1 \rangle 
\end{equation}
and in this form it impinges on the PBS$_2$.

Now we examine the evolution of the second two terms in Eq. (\ref{inputstate}) corresponding to the propagation of the signal qubit by the upper arm \textcircled{2}: In this case the control qubit enters the c-phase gate alone rendering the investigated terms to
$$
\beta \cos{\theta} |V_2 H_c \rangle + {\beta \mathrm{e}^{\imath \vartheta} \sin{\theta} \over \sqrt{3}} |V_2 V_c \rangle.
$$
Again, we post-select only on those cases, where the control qubit leaves the c-phase gate by the mode $c$. This post-selection is again assured by the QND gate which witnesses the control qubit presence with success probability of $\frac{1}{2}$. Considering also the action of HWP we get
$$
{\beta \cos{\theta} \over 2} |V_2 \rangle (|H_c \rangle + |V_c \rangle)  + {\beta \mathrm{e}^{\imath \vartheta} \sin{\theta} \over 2 \sqrt{3}} |V_2 \rangle (|H_c \rangle - |V_c \rangle).
$$
Subsequently the signal qubit undergoes a Hadamard transformation yielding
\begin{eqnarray*}
&& {\beta \cos{\theta} \over 2\sqrt{2}} \left( |H_2 H_c \rangle + |H_2 V_c \rangle - |V_2 H_c \rangle - |V_2 V_c \rangle \right) \\
&+& {\beta \mathrm{e}^{\imath \vartheta} \sin{\theta} \over 2 \sqrt{6}} \left( |H_2 H_c \rangle - |H_2 V_c \rangle - |V_2 H_c \rangle + |V_2 V_c \rangle \right).
\end{eqnarray*}
In the next step, both the signal and control qubits meet on the PPG gate's polarizing beam splitter. Since we post-select only on the cases where there is exactly one photon at the output of mode $c$, we continue considering only the terms
$$
{\beta \cos{\theta} \over 2\sqrt{2}} \left( |H_2 H_c \rangle - |V_2 V_c \rangle \right)
+{\beta \mathrm{e}^{\imath \vartheta} \sin{\theta} \over 2 \sqrt{6}} \left( |H_2 H_c \rangle + |V_2 V_c \rangle \right)$$
As usual, the control photon impinge on the detector. Depending on the outcome of the polarizing detection measurement performed, the signal qubit collapses into
\begin{equation}
\label{eq:signal_mode2}
{\beta \cos{\theta} \over 4} (|H_2 \rangle - |V_2 \rangle ) + {\beta \mathrm{e}^{\imath \vartheta} \sin{\theta} \over 4 \sqrt{3}} (|H_2 \rangle + |V_2 \rangle ) ,
\end{equation}
when diagonally polarized control qubit was detected or
$$
{\beta \cos{\theta} \over 4} (|H_2 \rangle + |V_2 \rangle ) +  {\beta \mathrm{e}^{\imath \vartheta} \sin{\theta} \over 4 \sqrt{3}} (|H_2 \rangle - |V_2  \rangle ) ,
$$
when we observe anti-diagonally polarized control photon. In the later case, we do apply a feed-forward consisting of a HWP placed at 0° causing $|V_2\rangle \rightarrow -|V_2\rangle$ and thus correcting the signal to the form of (\ref{eq:signal_mode2}). Subsequent Hadamard gate renders the signal to the form of
$$
{\beta \mathrm{e}^{\imath \vartheta} \sin{\theta} \over 2 \sqrt{3}} |H_2 \rangle +
{\beta \cos{\theta} \over 2} |V_2 \rangle.
$$
Before the signal impinges on the second polarizing beam splitter PBS$_2$, we introduce filtering by neutral density filter NDF of transmissivity $T = \frac{1}{2}$ to balance the amplitude with respect to the lower arm contribution described in (\ref{eq:arm1}).

After having both the terms of Eq. (\ref{inputstate}) evaluated, let us remind that the total state of the signal qubit reads
\begin{eqnarray*}
|\Psi_s\rangle = &&
{\alpha \cos{\theta} \over 2 \sqrt{2}} |H_1 \rangle +
{\alpha \mathrm{e}^{\imath \vartheta} \sin{\theta} \over 2 \sqrt{6}} |V_1 \rangle \\
+ &&{\beta \mathrm{e}^{\imath \vartheta} \sin{\theta} \over 2 \sqrt{6}} |H_2 \rangle +
{\beta \cos{\theta} \over 2 \sqrt{2}} |V_2 \rangle
\end{eqnarray*}
and after being subjected to the PBS$_2$ it takes the form of
$$
|\Psi_s\rangle = 
{\cos{\theta} \over 2 \sqrt{2}} (\alpha |H_1 \rangle - \beta |V_1 \rangle) +
{\mathrm{e}^{\imath \vartheta} \sin{\theta} \over 2 \sqrt{6}} ( \beta |H_2 \rangle + \alpha |V_2 \rangle ).
$$
Additional half-wave plate at 0° in first output corrects $-V \rightarrow + V$ and 
polarization swap $H\leftrightarrow V$ is applied to the second output by inserting a half wave plate at 45° yielding the final form of the signal state at the output of the router
\begin{equation}
|\Psi_s\rangle_\mathrm{OUT} = A_1 |\Psi_s \rangle_1 + A_2 |\Psi_s \rangle_2 ,
\end{equation}
where one can clearly observe the routing operation. The polarization state remains in the original form of (\ref{signalstate}), but the spatial degree of freedom is modified depending on the parameter $\theta$ of the control qubit. The ratio between the amplitudes $A_2$ and $A_1$ depend straightforwardly on this parameter
\begin{equation}
\tan{\chi} = {|A_2| \over |A_1|} = {\tan{\theta} \over \sqrt{3}},
\end{equation}
where we have introduced the routing ratio parameter $\chi$ in a similar manner as splitting ratio parametrization is introduced for ordinary beam splitters. Since $\theta$ lies in the interval $[0; {\pi\over 2}]$, the router can direct all the signal to the first output ($A_1$), to the second output ($A_2$) or to any of their superposition.

The success probability $P_\mathrm{succ} = |A_1|^2 + |A_2|^2$ does not depend on the signal state parameters $\alpha$ and $\beta$, but only on the ratio $\chi$ and therefore on the control qubit parameter $\theta$. One can easily find the relation between success probability and the routing parameter $\chi$ or the control qubit parameter $\theta$
\begin{equation}
P_\mathrm{succ} = \frac{1+2\cos^2\theta}{24} =  \frac{1+\tan^2\chi}{8+24\tan^2\chi}.
\end{equation}

\begin{figure}
\centerline{\includegraphics[scale=1]{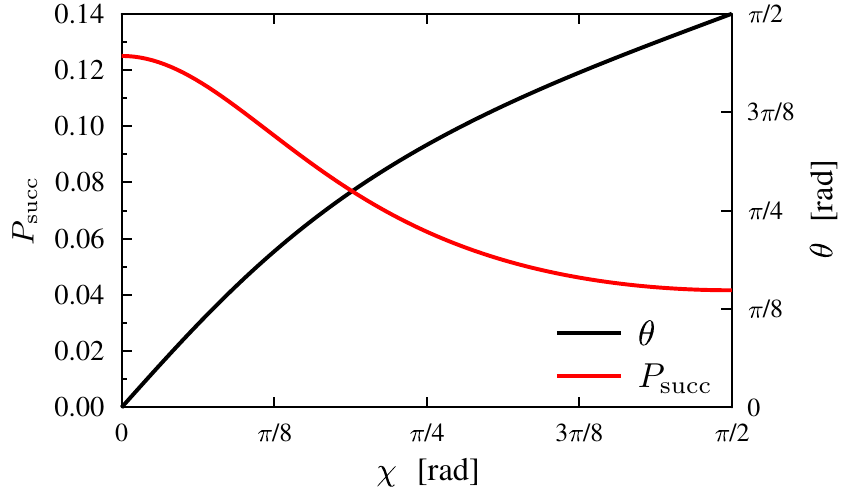}}
\caption{(Color online) Success probability of the routing procedure as a function of the routing ratio parameter $\chi$. Additionally the plot shows the relation between the routing parameter $\chi$ and the control qubit parameter $\theta$.}
\label{fig:psucc}
\end{figure}

It reaches maximum of $\frac{1}{8}$ for the $\chi = \theta = 0$ and, on the other hand, it is minimized for $\chi = \theta = \frac{\pi}{2}$ to a value of $\frac{1}{24}$. The plot in Fig. \ref{fig:psucc} shows success probability of routing as a function of the routing parameter $\chi$. It also depicts the relation between the routing parameter $\chi$ and the control qubit parameter $\theta$.

Note that neutral density filter of intensity transmissivity of $\frac{1}{3}$ can be placed to output port $s_1$ in order to equalize the success probability so it is completely control state independent and fixed to the value of $\frac{1}{24}$.

\section{Tunable c-phase gate based router}
\begin{figure}
\includegraphics[scale=0.8]{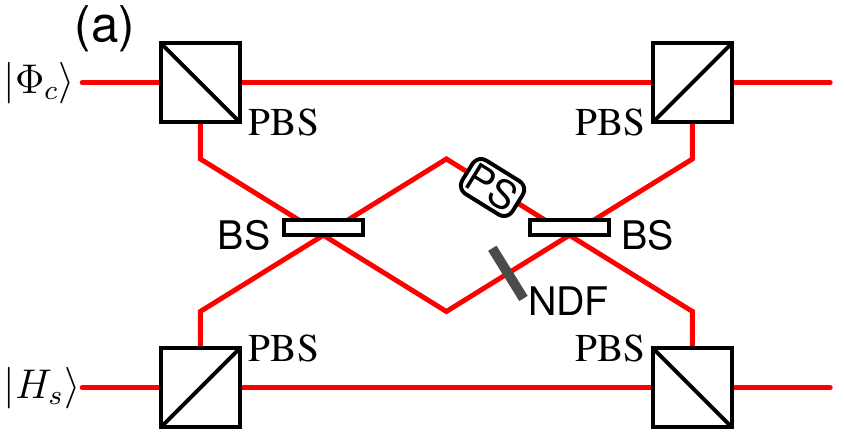}
\includegraphics[scale=0.8]{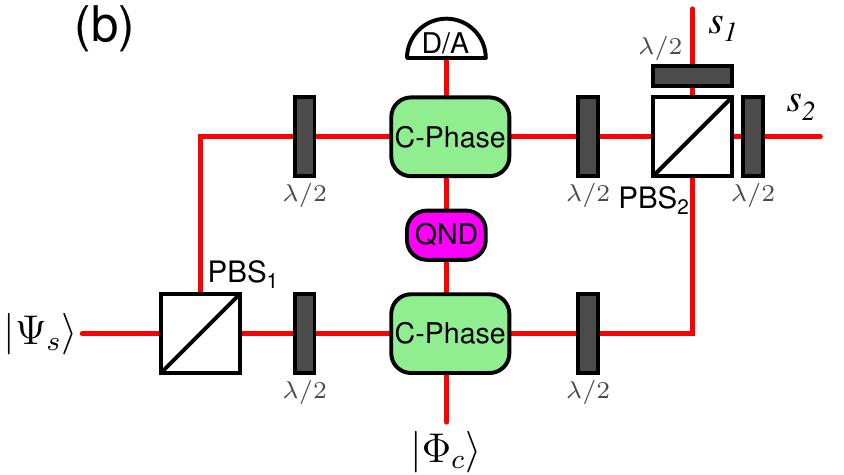}
\caption{\label{fig:tunable_cphase} (Color online) (a) Linear-optical scheme of a tunable c-phase gate. The interaction Mach-zehnder interferometer replaces the fixed polarization dependent beam splitter PDBS in the original scheme. The phase shift imposed by the gate is tuned by setting specific values of phase shift (PS) and losses (NDF) in this interferometer (for more details see Ref. \cite{Lemr_PRL106}). (b) Modified scheme of the router using two tunable c-phase gates.}
\end{figure}
In the previous section we have only considered c-phase gate with a fixed phase shift of $\varphi = \pi$ (also known as the controlled-sign gate). In some cases, however, higher success probability can be achieved using a tunable c-phase gate that can be set to exercise a phase shift $\varphi$ of any value in the range $[0;\pi]$. The reason for considering this tunable c-phase gate lies in the fact that the success probability of the gate is a function of its phase shift. In 2010, Konrad Kieling and his colleagues \cite{Kieling10} discovered the success probability $P_C$ relation to the phase shift $\varphi$ \footnote{$P_C$ is the optimal success probability achievable using only linear optics and vacuum ancillae.}
\begin{eqnarray}
\label{eq:cphase_psucc}
   P_\mathrm{C}= \left[1+2\left|\sin\frac{\varphi}{2}\right|+2^{3/2}\sin\left(\frac{\pi-\varphi}{4}\right){\left|\sin\frac{\varphi}{2}\right|^{1/2}}\right]^{-2}.
\end{eqnarray}
In order to make the c-phase gate tunable, one needs to replace the fixed polarization dependent beam splitter by a interaction Mach-Zehnder interferometer with tunable phase and losses (see Fig. \ref{fig:tunable_cphase}a).

In a recent paper \cite{lemr12prob}, we showed that if one does not seek this success probability to be input state independent, the c-phase gate can be generalized to perform the transformation
\begin{eqnarray}
\label{eq:cphase_general}
|H_1H_c\rangle &\rightarrow & |H_1H_c\rangle \nonumber \\
|H_1V_c\rangle &\rightarrow & \sqrt{A_C}|H_1V_c\rangle \nonumber \\
|V_1H_c\rangle &\rightarrow & \sqrt{A_C}|V_1H_c\rangle \nonumber \\
|V_1V_c\rangle &\rightarrow & A_C \mathrm{e}^{\imath \varphi} |V_1V_c\rangle \nonumber \\
|V_2H_c\rangle &\rightarrow & |V_2H_c\rangle  \nonumber \\
|V_2V_c\rangle &\rightarrow & \sqrt{A_C}|V_2V_c\rangle,
\end{eqnarray}
where we have already adopted the notation of Eq. (\ref{inputstate}) and introduced $|A_C|^2 = P_C$. Note that in the first and fifth case, there is no photon entering the interaction interferometer, while in the second, third and sixth case there is exactly one photon entering this interferometer. In the fourth case, both the photons are subjected to the interferometer. The success probability thus becomes input state dependent, but higher in average.

When using the tunable c-phase gate for routing, several modifications of the setup have to be put in place (see Fig. \ref{fig:tunable_cphase}b): Firstly, the PPG gate  has to be replaced by a second c-phase gate set to the same phase shift as the first one. 
This c-phase gate would exercise the same transformation as described by (\ref{eq:cphase_general}), but with indexes $1 \leftrightarrow 2$ swapped.
Secondly, the transformation (HWP) performed on the control qubit between the two interaction gates is removed, so is the NDF in upper arm \textcircled{2}. And finally, a generalized transformation in both arms \textcircled{1} and \textcircled{2} before the signal enters the c-phase gates has to be recalculated. This transformation reflecting the success probability of the c-phase gate reads
$$
|H_1\rangle \rightarrow \frac{\sqrt{A_C}}{\sqrt{1+A_C}}|H_1\rangle + \frac{1}{\sqrt{1+A_C}}|V_1\rangle
$$
in \textcircled{1} and
$$
|V_2\rangle \rightarrow \frac{\sqrt{A_C}}{\sqrt{1+A_C}}|H_2\rangle - \frac{1}{\sqrt{1+A_C}}|V_2\rangle
$$
in \textcircled{2}. This transformation assures the signal state independence of the routing procedure.

After incorporating these modifications, one can proceed in exactly the same manner as in Section II to reveal that the signal state just before impinging on the PBS$_2$ takes the form of
\begin{eqnarray}
\label{eq:tunable_arm1}
&& \frac{\alpha\sqrt{A_C}}{2\sqrt{2+2A_C}} \left[ 2 \cos\theta +A_C \mathrm{e}^{\imath \vartheta} \sin\theta (1+\mathrm{e}^{\imath \varphi}) \right] |H_1\rangle \nonumber \\
&& + \frac{\alpha \mathrm{e}^{\imath \vartheta} \sin\theta\sqrt{A^3_C}}{2\sqrt{2+2A_C}}(1-\mathrm{e}^{\imath \varphi})|V_1\rangle
\end{eqnarray}
in the first arm \textcircled{1} and the form of
\begin{eqnarray}
\label{eq:tunable_arm2}
&& \frac{\beta\sqrt{A_C}}{2\sqrt{2+2A_C}} \left[ 2 \cos\theta + A_C \mathrm{e}^{\imath \vartheta} \sin\theta (1+\mathrm{e}^{\imath \varphi})\right]|V_2\rangle \nonumber \\
&& + \frac{\beta \mathrm{e}^{\imath \vartheta} \sin\theta\sqrt{A^3_C}}{2\sqrt{2+2A_C}}(1-\mathrm{e}^{\imath \varphi})|H_2\rangle
\end{eqnarray}
in the second arm \textcircled{2}. The Eqs. (\ref{eq:tunable_arm1}) and (\ref{eq:tunable_arm2}) indicate that the value of the phase shift $\varphi$ imposed by the gate limits the routing ratio $\chi$
$$
\tan \chi = \frac{A_C\sin\theta|1-\mathrm{e}^{\imath \varphi}|}{|2 \cos\theta + A_C \mathrm{e}^{\imath \vartheta} \sin\theta (1+\mathrm{e}^{\imath \varphi})|}.
$$
For the sake of readability, we limit to $\vartheta = 0$ in the remaining part of this section. Thus even for $\theta = \frac{\pi}{2}$ the routing ratio is bound by the relation
\begin{equation}
\label{eq:routing_limit}
\chi_L(\varphi) \equiv \mathrm{max}\lbrace\chi(\varphi,\theta)\rbrace_\theta = \mathrm{atan} \left(\!\frac{1-\cos\varphi}{\sin\varphi}\!\right) = \frac{\varphi}{2},
\end{equation}
where we have introduced the routing ratio limit $\chi_L$ for a given phase shift $\varphi$.

Using this definition, we can formulate the range of achievable routing ratios for a given phase shift $\varphi$ to be $\chi\in[0;\chi_L]$ obtained when tuning monotonously the control qubit in the range $\theta\in[0;\frac{\pi}{2}]$. Note that, as expected, $\chi_L = 0$ for $\varphi = 0$ and $\chi_L = \frac{\pi}{2}$ for $\varphi = \pi$.

\begin{figure}
\includegraphics[scale=1]{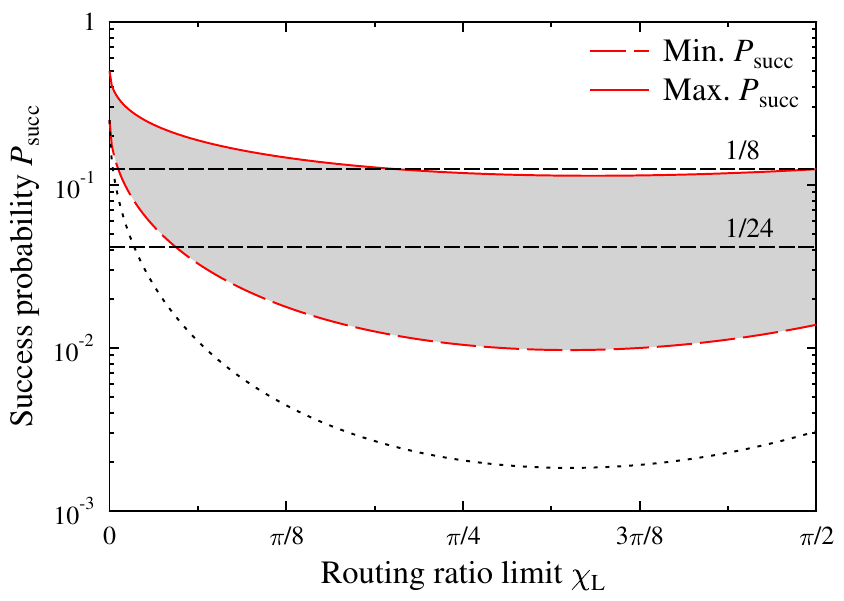}
\caption{\label{fig:tune_prob} (Color online) Maximum and minimum success probability of the router using two tunable c-phase gates depicted as functions of the routing ratio limit. The plot indicates that for sufficiently small values of routing ratio limit, the tunable c-phase gate offers better performance then the scheme devised in previous section (maximum and minimum success probability of the previous scheme is depicted using black dashed lines at $\frac{1}{8}$ and $\frac{1}{24}$). The dotted line shows the success probability if standard state-independent c-phase gate is used instead of the generalized version.}
\end{figure}

We can also easily find the success probability of the router by calculating the norm of the output state (\ref{eq:tunable_arm1}) and (\ref{eq:tunable_arm2}). This success probability function depends on both $\varphi$ and $\theta$. For a fixed value of $\varphi$ we can find its minimum of
\begin{equation}
\label{eq:tunable_router_successMIN}
\mathrm{min}\lbrace P_\mathrm{succ}(\varphi,\theta)\rbrace_\theta = \frac{\sqrt[3]{A^4_C}}{2+2A_C}
\end{equation}
always for $\theta = \frac{\pi}{2}$. On the other hand, the maximum success probability is obtained for different values of $\theta$ depending on the phase shift $\varphi$
\begin{equation}
\label{eq:tunable_router_successMAX}
\theta|_{P_\mathrm{MAX}} = \frac{1}{2} \mathrm{atan}\left[\frac{A_C(1-\cos\varphi)}{1-A^2_C}\right].
\end{equation}

Both the success probability and the routing ratio limit are function of the gate phase shift $\varphi$. To illustrate their mutual relation, we have plotted the success probability as a function routing limit in Fig. \ref{fig:tune_prob}. In this figure, we show the maximum and minimum success probability together with the range between them for a given routing ratio limit. Observing this plot, we conclude that for sufficiently small values of routing ratio limit (e.g. $\chi_L = \frac{\pi}{8}$) the tunable c-phase gate offers increased success probability in comparison with the scheme proposed in previous section. On the other hand, for larger routing ratio limits (closer to $\frac{\pi}{2}$), the former scheme does better since it uses only one c-phase gate together with a more efficient PPG gate.

\section{Multi-qubit routing}
\begin{figure}
\includegraphics[scale=0.8]{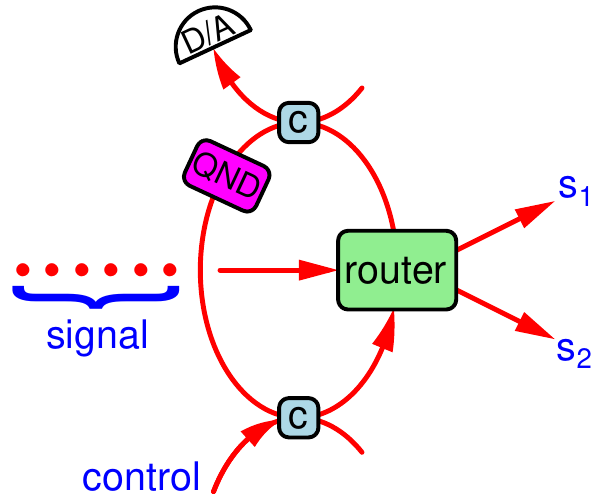}
\caption{\label{fig:concept2} (Color online) Scheme for multi-qubit routing, c denotes classical optical coupler.}
\end{figure}
So far we have only considered routing a single signal qubit using one control qubit. In principle, however, the device can route a chain of signal qubits making use of the same control qubit. The Fig. \ref{fig:concept2} depicts the configuration for such multi-qubit routing. In this model, two c-phase gates are used as in the previous section, but the control qubit is not detected immediately. After interacting with the first signal qubit, the control qubit presence is verified in a QND gate and then it is transferred back to the input of the router so it meets with the second signal qubit. After interacting with the last signal qubit, the control qubit is subjected to polarization analysis thus projecting the state of all signal qubits.

Let us assume the state of $n$-th signal qubit to be
$$
|\Psi_n\rangle = \alpha_n|H\rangle + \beta_n|V\rangle.
$$
Thus the state of the entire system can be expressed as
$$
|\Psi_c\rangle\prod_n|\Psi_n\rangle_\mathrm{IN} = (\cos\theta |H\rangle + \mathrm{e}^{\imath \vartheta}\sin\theta |V\rangle)\prod_n|\Psi_n\rangle_\mathrm{IN},
$$
where the control qubit $|\Psi_c\rangle$ has been introduced as in previous sections. The procedure described in the previous paragraph then renders the state to
$$
\cos\theta |H\rangle \prod_n|\Psi_n\rangle_\mathrm{OUT1} + \frac{\mathrm{e}^{\imath \vartheta}\sin\theta}{3} |V\rangle \prod_n|\Psi_n\rangle_\mathrm{OUT2},
$$
where OUT1 and OUT2 denote output ports of the router. Subsequent detection of the control qubit in $|H\rangle\pm|V\rangle$ basis projects the state to the required form of
$$
\cos\theta \prod_n|\Psi_n\rangle_\mathrm{OUT1} + \frac{\mathrm{e}^{\imath \vartheta}\sin\theta}{3} \prod_n|\Psi_n\rangle_\mathrm{OUT2}.
$$
Since the routing operation is probabilistic, the success probability of routing $n$ qubits decreases exponentially with $n$. Considering the success probability of one run of the router to be $P_\mathrm{succ}$ (as calculated in previous section), the $n$-qubit router will perform with success probability
$$
P_\mathrm{total} = 2^{1-4n}\left(1-\frac{8}{9}\sin^2\theta\right)^n.
$$

It is worth noting that, apart from routing, this procedure can be used to generate the so-called noon states \cite{bib:fiurasek:nphoton,bib:mitchell:super}. These states of the form of $(|N0\rangle + |0N\rangle)$, where $N$ denotes the number of photons, are useful for instance in quantum metrology \cite{nagata07noon} or quantum litography \cite{bib:bjork:litography}.

\section{Conclusions}
We have presented an all-linear-optical scheme for a fully quantum router. The router meets all the requirements presented in the introduction. This concept is more practical then some of the previously presented routers making use of experimentally difficult light-matter interaction or unrealistic strengths of non-linear optical phenomena \cite{Zueco09,Aoki09,Hoi11,Hall11}. In contrast to previously published linear-optical schemes, our router provides genuine quantum routing being controlled by a qubit while not disturbing the state of the signal qubit (in contrast to \cite{chang12router}). Also it only requires one control qubit to route a single signal qubit making it more resource-efficient (in contrast to \cite{Lemr12_router}). 

The proposed scheme operates with success probabilities ranging between $\frac{1}{8}$ and $\frac{1}{24}$ depending on the control qubit state. We also discuss optimization of the success probability that can be reached by using tunable c-phase gates. The presented analysis shows how efficiency can be increased upto $\frac{1}{2}$ if only small amplitude of the signal is needed to be sent to output port 2. Note also that detector efficiency just scales the success probability, but does not change fidelity of the output.

We have also provided a recipe for multi-qubit router, where the same control qubit is used to route general number of signal qubits. In this case, the success probability of the procedure scales exponentially with the number of routed signal qubits.

\section{Acknowledgement}
The authors gratefully acknowledge the support by the Operational Program Research and Development for Innovations -- European Regional Development Fund (project No. CZ.1.05/2.1.00/03.0058. KL acknowledges the support by the Czech Science Foundation (project No. 13-31000P), KB acknowledges project No. CZ.1.07/2.3.00/30.0041 by the Czech Ministry of Education, A\v{C} acknowledges project No. CZ.1.07/2.3.00/20.0017 by the Czech Ministry of Education and JS acknowledges support by the project AVOZ10100522 by the Institute of Physics of CAS.

\end{document}